\documentclass[conference]{IEEEtran}

\usepackage{url}
\usepackage[pdftex]{graphicx}

\usepackage{balance} 
\usepackage{booktabs}
\usepackage{siunitx}
\usepackage{biblatex} 

\usepackage{amsmath}
\usepackage{cleveref}

\usepackage{fancyvrb}
\usepackage[most]{tcolorbox}
\usepackage[colorinlistoftodos,prependcaption,textsize=tiny]{todonotes}

\bibliography{references.bib}

\title{How is Google using AI for internal code migrations?}

\author{\IEEEauthorblockN{
Stoyan Nikolov\IEEEauthorrefmark{1}, Daniele Codecasa\IEEEauthorrefmark{1}, Anna Sjövall\IEEEauthorrefmark{1}, Maxim Tabachnyk\IEEEauthorrefmark{1}, Satish Chandra\IEEEauthorrefmark{1},\\ Siddharth Taneja\IEEEauthorrefmark{2}, Celal Ziftci\IEEEauthorrefmark{2}
}
\IEEEauthorblockA{
\IEEEauthorrefmark{1}{Google Core}, \IEEEauthorrefmark{2}{Google Ads}\\
Email:\{stoyannk, cdcs, annaps, tabachnyk, chandrasatish, tanejas, ziftci\}@google.com
}
}

\begin{document}

\maketitle

\begin{abstract}
In recent years, there has been a tremendous interest in using generative AI, and particularly large language models (LLMs) in software engineering;  indeed there are now several commercially available tools, and many large companies also have created proprietary ML-based tools for their own software engineers. While the use of ML for common tasks such as code completion is available in commodity tools, there is a growing interest in application of LLMs for more bespoke purposes.   One such purpose is code migration.

This article is an experience report on using LLMs for code migrations at Google. It is not a research study, in the sense that we do not carry out comparisons against other approaches or evaluate research questions/hypotheses.  Rather, we share our experiences in applying LLM-based code migration in an enterprise context across a range of migration cases, in the hope that other industry practitioners will find our insights useful.  Many of these learnings apply to any application of ML in software engineering.  We see evidence that the use of LLMs can reduce the time needed for migrations significantly, and can reduce barriers to get started and complete migration programs.
\end{abstract}

\section{Introduction}\label{sec:Introduction}
Google Product Areas (PAs) such as Ads, Search, Workspace and YouTube are mature software development organizations that perhaps resemble many Fortune 500 companies in terms of software development challenges:

\begin{itemize}
  \item Mature (20+ years), large code bases.
  \item Need to keep up with business demands for agility in a competitive external environment.
  \item Need to maintain code and use new frameworks etc. to ascertain current feature demands.
  \item Software development quality and agility are differentiators in the market, even though what these PAs offer in their markets is not a software product as such.
\end{itemize}

Over the years, Google as a whole has adopted software engineering principles – monorepo, analysis tools, a rigorous code review, CI/CD etc. – that have served all PAs well. See \cite{winters2020software}.

In the era of LLMs, the practice of software engineering is going through an industry-wide transformation.  
%Already, technologies such as Copilot and chatGPT have made significant strides in the market and promised a new set of tools for developer productivity. \cite{peng2023impactaideveloperproductivity,web:codewhsiperer}. 
%New innovations in UX of AI helping developers is making rapid progress, e.g. Cursor or Replit \cite{web:replit,web:cursor}.
%How has Google started to use these new ML-based technologies internally in software engineering, and what has our experience been so far? 
This profound change is taking place at Google as well. Google uses AI technologies in software engineering internally at two levels:

First is the \textbf{generic} AI-based tooling for software development that is designed for all Googlers across all PAs.  These technologies – built by Google for Google – include code completion, code review, question answering, and so on.
The generic AI-based development tools have been very successful at Google (see Section~\ref{ai-techs}) as well as in the external community, thanks to the work of Github and other companies \cite{peng2023impactaideveloperproductivity,dunay2024multilineaiassistedcodeauthoring}. However, by necessity the IDE-based tooling is designed for “mass market” use, where the UX is paramount and typically, the value added per interaction is small but it happens many times (e.g. code completion).  This is the space the mass market tools / generic tools optimize for. We have previously talked about this in blog posts and papers \cite{web:blog:aigooglecc,web:blog:aigoogle,C2CPaper}.

Second is \textbf{bespoke} solutions for each PA.  Examples include specific code migrations, code efficiency optimization, and test generation tasks, where we have effectively used LLMs in custom (or ``bespoke'') ways.  
As opposed to mass market, or generic tools, the number of interactions may be smaller for bespoke tools, but the complexity of each interaction is often higher. Furthermore, the emphasis here is often to accomplish an end-to-end task rather than just being an IDE convenience. For instance, in code migration at scale, the need is to be able to perform repo-level change correctly and consistently.  These need custom solutions.  They might use the same primitives as mass market tools, but the goals are different.

This article focuses on such bespoke solutions, specifically on the migration workloads that we see from Google product units.  We describe the setting of migration projects in Google PAs, and the challenges we address in making LLM-based code migrations deliver the intended business value.  The measure of success we adopt is whether there is \emph{at least a 50\% acceleration} in task completion rate in an ongoing project  (see Section~\ref{migration-techs}.)

Section \ref{ads} in the paper covers a case study from the Google Ads PA.  The Google Ads business is one of the largest in the world and is built on a code base of 500+M lines of code.
Ads code base uses several IDs that were 32 bits, but need to be converted to 64 bits to avoid negative rollover scenarios that could cause outages.
With the use of the LLM-based approach (see details in Section~\ref{ads}), we are on track to achieve our migration targets, surpassing the success metric of 50\% or better acceleration.

Building on the success of the Ads experience, the number of different migration initiatives around Google has been increasing.
In the paper, we discuss three other distinct migrations problems arising from different Google PAs:
\begin{itemize}
\item JUnit3 to JUnit4 migration (Section~\ref{sec:j3})
\item Joda time to Java time migration (Section~\ref{sec:goodtime})
\item Cleanup of experimental flags (Section~\ref{sec:mendel})
\end{itemize}

\begin{figure}[h]
\centering
\includegraphics[width=1\linewidth]{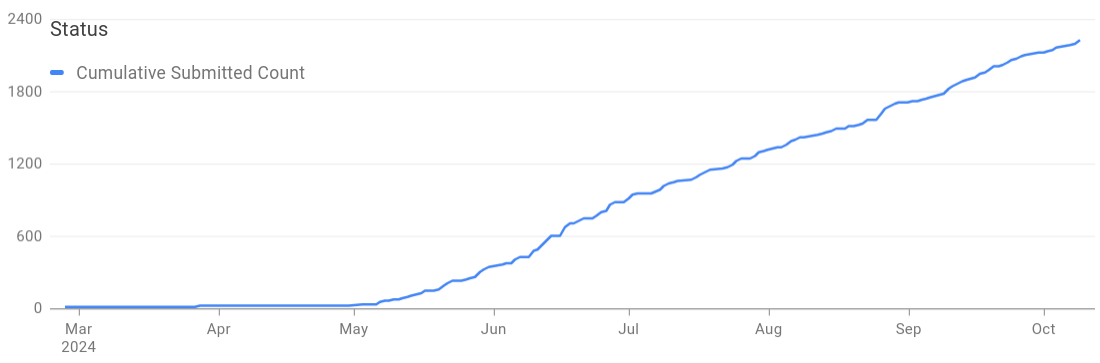}
\caption{Landed changelists of AI-powered migrations for the first 3 quarters for 2024.}
\label{fig:cls}
\end{figure}

While the above are a good representative set, we
have worked on several additional migrations: in fact, 
through 2024 the number of changelists (similar to pull requests in Git) from migration efforts across Ads and other product areas has steadily increased (see Fig~\ref{fig:cls}) and the types of changes supported has expanded significantly. This has created an ecosystem that allows the teams to employ economies of scale, slotting new migrations into already proven workflows.

Not only did the use of LLMs accelerate these migrations, from an organizational viewpoint, we have been able to complete complex migrations that were stalled for several years and required continued attention from the business.  We have completed efforts that spanned several teams using a handful of engineers and saved the business hundreds of engineers worth of work.

Achieving success in LLM-based code migration is not straightforward.  The use of LLMs alone through simple prompting is not sufficient for anything but the simplest of migrations. Instead, as we found through our journeys, and as described in the case studies in this paper, \textit{a combination of AST-based techniques, heuristics, and LLMs} are needed to achieve success.  Moreover, rolling out the changes in a safe way to avoid costly regressions is also important. 

Although each migration is different, and requires bespoke work, they often follow similar patterns.  The cases described in this paper show a set of such patterns that, we believe, will continue to appear in future migration projects.  We believe that the techniques described are not Google-specific and we expect that they can be applied to any LLM-powered code migration at large enterprises.

To help with the challenges of migrations and to leverage the commonality that we see across the many cases, we have developed a common toolkit (Sec ~\ref{sec:toolkit}) that we used for the code changes and the techniques to find the relevant files to change.  The project-specific customization comes from the LLM prompts used, as well as in the validation steps for the code changes and the review and rollout phases, which are still largely human-driven.

%\todo{a summary of takeaways ... why shouldanyone read this paper.}

The rest of the paper is organized as follows: Section \ref{ai-techs} briefly recaps the generic code AI technologies that we use at Google.  
Section \ref{migration-techs} talks about the bespoke technologies, focusing on code migration. Following this section, there are four sections, each describing a different code migration case study.  Finally, 
Section \ref{learnings} discusses our learning and takeaways from our experience.

\section{Generic AI tools in Google internal software development}\label{ai-techs}

Ever since the advent of powerful transformer-based models, we started exploring how to apply LLMs to software development. LLM-based inline code completion is the most popular application of AI applied to software development: it is a natural application of LLM technology to use the code itself as training data. The UX feels natural to developers since word-level autocomplete has been a core feature of IDEs for many years. Also, it’s possible to use a rough measure of impact, e.g., the percentage of new characters written by AI. For these reasons and more, it made sense for this application of LLMs to be the first to deploy.

We have seen continued fast growth similar to other enterprise contexts \cite{dunay2024multilineaiassistedcodeauthoring}, with an acceptance rate by software engineers of 38\% assisting in the completion of 67\% of code characters \cite{web:blog:aigoogle}, defined as the number of accepted characters from AI-based suggestions divided by the sum of manually typed characters and accepted characters from AI-based suggestions. In other words, the amount of characters in the code that are completed with AI-based assistance is now higher than manually typed by developers. 
While developers still need to spend time reviewing suggestions, they have more time to focus on code design.

Key improvements came from both the models — larger models with improved coding capabilities, heuristics for constructing the context provided to the model, as well as tuning models on usage logs containing acceptances, rejections and corrections — and the UX. This cycle is essential for learning from practical behavior, rather than synthetic formulations.

\begin{figure}[h]
\centering
\includegraphics[width=0.5\linewidth]{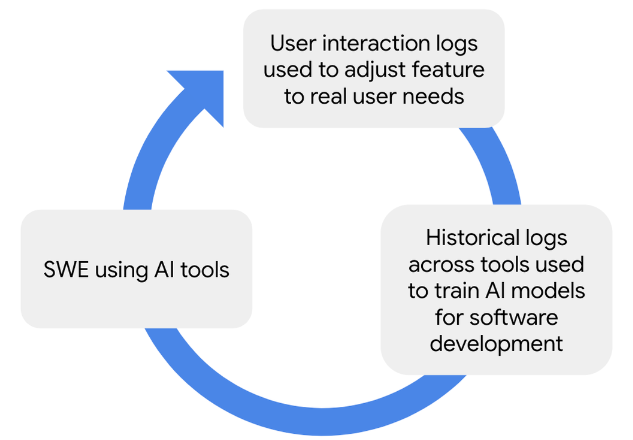}
\caption{Improving AI-based features in coding tools (e.g., in the IDE) with historical high quality data across tools and with usage data capturing user preferences and needs.}
\label{fig:wheel}
\end{figure}

We use our extensive and high quality logs of internal software engineering activities across multiple tools, which we have curated over many years. This data, for example, enables us to represent fine-grained code edits, build outcomes, edits to resolve build issues, code copy-paste actions, fixes of pasted code \cite{web:blog:spaste}, code reviews, edits to fix reviewer issues, and change submissions to a repository. The training data is an aligned corpus of code with task-specific annotations in input as well as in output. The design of the data collection process, the shape of the training data, and the model that is trained on this data was described in our DIDACT \cite{web:blog:didact} blog. We continue to explore these powerful datasets with newer generations of foundation models available to us (discussed more below).

Our next significant deployments were resolving code review comments \cite{web:blog:c2c} (\textgreater8\% of which are now addressed with AI-based assistance) and automatically adapting pasted code \cite{web:blog:spaste} to the surrounding context (now responsible for $\sim$2\% of code in the IDE). Further deployments include instructing the IDE to perform code edits with natural language and predicting fixes to build failures \cite{web:blog:build}. Other applications, e.g., predicting tips for code readability \cite{Vijayvergiya_2024} following a similar pattern are also possible.

Together, these deployed applications have been successful, highly-used applications at Google, with measurable impact on productivity in a real, industrial context.

\begin{figure}[h]
\centering
\includegraphics[width=1\linewidth]{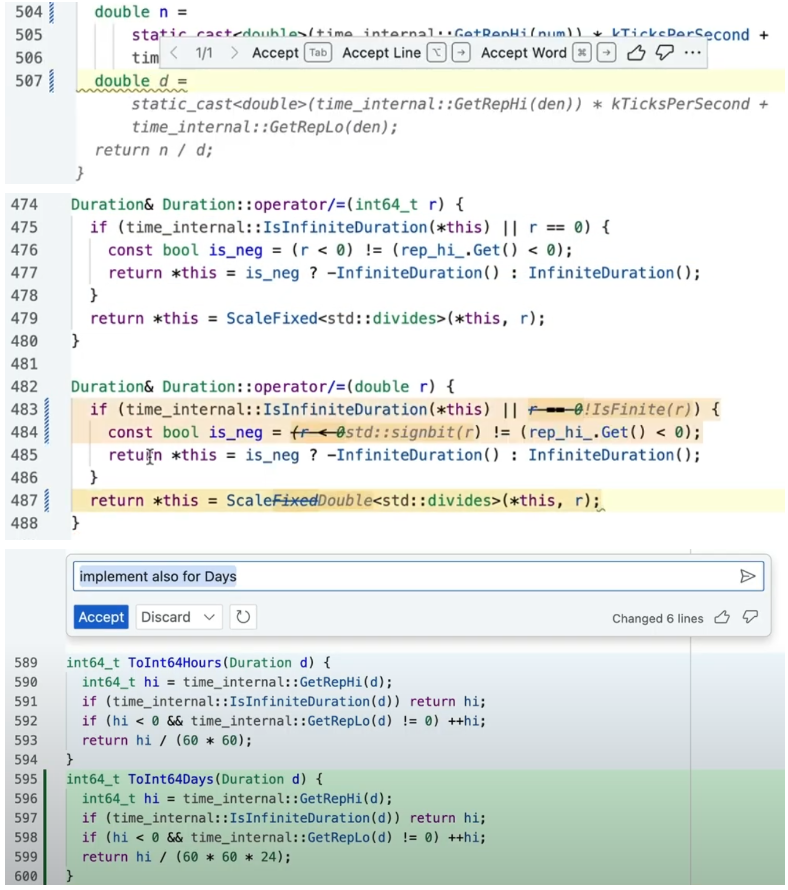}
\caption{A demonstration of how a variety of AI-based features can work together to assist with coding in the IDE. top: code completion, middle: adjusting copy-pasted code to the context, bottom: code edits based on natural language instructions. See our blog for more details \cite{web:blog:aigoogle}.}
\label{fig:code_demostrations}
\end{figure}

\section{Bespoke use of LLMs for Code Migration}\label{migration-techs}
In this section we describe some of the LLM-powered efforts that Google implemented to address lingering technical debt and code migrations within several PAs.  

The need for code migration is not new at Google (and neither elsewhere.) At Google’s monorepo scale, special infrastructure for large-scale changes \cite{winters2020software} was, and still is used. This has allowed huge migrations like programming language version changes, API deprecations etc. at a fraction of the cost of doing them manually. That infrastructure uses static analysis and tools like Kythe \cite{web:blog:kythe}, Code Search \cite{winters2020software} and ClangMR \cite{web:blog:clang}.

However, for the kinds of code migrations that we wish to accomplish, these ``deterministic'' code change solutions have not proven to be quite as effective. This is because the contextual clues and the actual changes to be made have quite a bit of variance, and these are difficult to write out in a deterministic code transformation pass.  This is exactly where modern LLMs are very effective, and in practice, offer a lower barrier to entry compared to devising other customized program transformation systems, such as based on program synthesis.  Our goal is to find opportunities where LLMs will provide additional value by not requiring hard to maintain AST (abstract syntax trees)-based transformations, and at the same time, have scale of application that justifies the work. While AST-based techniques offer deterministic change generation, the cases we want to tackle can span complex code constructs that would be hard to implement as ASTs to cover all cases.

At a high level, the use of an LLM prompt to make a task-specific code change might look like the following:

\begin{scriptsize}
\begin{verbatim}
    You are a software engineer tasked to do {TASK}.
    Here are the rules to accomplish this:
    1. First rule
    2. Second rule
    ...
\end{verbatim}
\end{scriptsize}

Where \texttt{\scriptsize rule} refers to an informal description of what needs to be accomplished; as opposed to a precise AST-level pattern matching and transformation.

We use LLM prompting to build common workflows that contain bespoke, customised parts---per-task instructions to the model itself---as well as some of the sub-steps in the code migration like the file discovery and validations. This  approach allows us to quickly onboard new use cases, because we built a palette of reusable sub-steps which we can combine and adapt.

\begin{figure*}[h]
\centering
\includegraphics[scale=0.45]{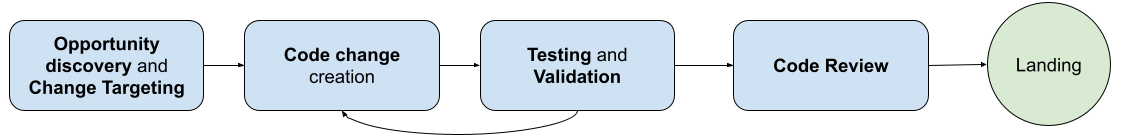}
\caption{The high-level process to land an AI-authored change in the monorepo.
We use LLMs extensively in code change creation, and partly in discovery and validation phases.}
\label{fig:mono}
\end{figure*}

We emphasize that LLMs are only one part of the complete solution, which includes some AST and symbol-based techniques, and some purely process issues such as change rollout.  The LLM role is focused on the edit generation.  The parts where we need to identify locations at which to make changes, and where we need to validate that the right thing took place, are handled mostly using deterministic AST techniques with a few cases supported by the LLM as well.  In Fig~\ref{fig:mono} we provide a conceptual diagram of the step-by-step process. After the opportunities for the change are discovered, they are generated using the LLM and a loop begins that iterates on tests and other validations until the changes are deemed good. Humans review the LLM-generated code the same way as any other code and they add any missing tests to cover the changed lines. The final step of landing (executing them in a production environment) the changes depends on the product they are part of.

%\paragraph*{How to evaluate the success of these initiatives?}\label{success}

As mentioned in Section~\ref{sec:Introduction}, for each migration described we have defined success as AI \textit{saving at least 50\% of the time for the end-to-end work}. This time is not only the change generation itself (the code rewrite), but also finding the places to migrate, doing reviews and rolling out the changes.  

This is notably different from the success metric typically used in application of generic technologies (such as code completion, where we typically report percent of code written by AI, or the acceptance rate). We found that while the ratio of code generated by AI that gets committed is a good proxy for the time savings in most cases, it does not always capture all the value. Anecdotal remarks from developers suggest that even if the changes are not perfect, there is a lot of value in having an initial version of the changelist already created, through which developers can quickly find the places where the changes are needed.  Thus, we believe that success should be assessed based on time saving on the end-to-end work. The impact on code quality is also important.  Currently the quality is assured through the manual review process of the code - same as for purely human-generated code.
Whether there is a long-term impact on quality remains to be seen.

In all of our case studies, the speed-up provided by the AI on the code changes alone was higher, but we believe that anchoring the success metric on the whole development journey offers a clearer impact goal and is aligned with the business outcome, rapid completion of the migrations. For fully completed migrations, we were able to estimate accurately the time savings (mostly based on historical data), while for some of the ongoing ones (such as Joda time API migration, Section~\ref{sec:goodtime}) we relied on the expert engineers estimating a time-saving for a set of changelists created by the tools and we extrapolated from that.

\section{int32 to int64 ID migration}\label{ads}

We previously shared information about this work in this blog post \cite{web:blog:gmigrations}.

Google Ads has dozens of numerical unique “ID” types used as handles — for users, merchants, campaigns, etc. — and these IDs were originally defined as 32-bit integers in C++ and Java. But with the current growth in the number of IDs, we expect them to overflow the 32-bit capacity much sooner than originally expected.

Although some migration was done initially manually, we decided to employ an LLM-powered code migration flow, described below, to accelerate the work.
There are several challenges related to the code changes that needed addressing:
\begin{itemize}
    \item The IDs are often defined as generic numbers (\texttt{int32\_t} in C++ or \texttt{Integer} in Java) and are not of a unique, easily searchable type, which makes the process of finding them through static tooling non-trivial.
    \item There are tens of thousands of code locations across thousands of files where these IDs are used.
    \item Tracking the changes across all the involved teams would be very difficult if each team were to handle the migration in their data themselves.
    \item Changes in the class interfaces need to be taken into account across multiple files.
    \item Tests need to be updated to verify that the 64-bit IDs are handled correctly.
    \item Constants are sometimes expressed as C++ macros. IDs are occasionally serialized as text, away from where they are used.
\end{itemize}

The full effort, if done manually was expected to require hundreds of software engineering years and complex cross-team coordination. The approach Google has had for such cases is to have a central team that drives the migration.
It is what we did in this case as well and devised the following workflow:

\begin{enumerate}
    \item An expert engineer from Ads finds the ID they want to migrate and, using a combination of Code Search, Kythe~\cite{web:blog:kythe}, and custom scripts, identifies a (best effort) superset of files and locations to migrate.
    \item An LLM-based migration toolkit, triggered by an expert, runs autonomously and produces verified changes that only contain code that passes unit tests. When necessary, tests are also updated to reflect the changes to the code.
    \item The same engineer then manually checks the change and potentially updates files where the model failed or made a mistake. The changes are then sharded and sent to multiple reviewers who own the part of the codebase affected by the change.
\end{enumerate}

We found that 80\% of the code modifications in the landed CLs were fully AI-authored, and the rest were human-authored or edited from the AI suggestions.  We calculate the percentage of AI-authored code by tracking the state of the changelists (similar to pull requests in git). The first version (known internally as snapshot) of the changelist is the one generated by the LLM-powered tooling. The version actually committed to the repo is compared to this and a per-character difference is calculated.

We discovered that in most cases, the human needed to revert at least some changes the model made that were either incorrect or not necessary.  Given the complexity and sensitive nature of the modified code, effort has to be spent in carefully rolling out each change to users.  This observation led to further investment in LLM-driven verification (described later) to reduce this burden.

The total time spent on the migration was reduced by an estimated 50\% as reported by the engineers doing the migration, when compared to a similar exercise carried out without LLM assistance.  In this calculation, we take into account the time needed to review and land the changes.  There was also a significant reduction in communication overhead, as a single engineer could generate all necessary changes. 

We will now discuss some detail about each of the steps in the workflow for this migration:

\subsubsection{Finding code locations to modify}

In this migration, the main target is usually one or more fields defined in a protocol buffer\cite{web:protobuf} - a standard mechanism for serializing structured data. We use a variety of technique to identify the final files and code lines to modify.

First, we start with the manual identification of the protocol buffer fields for an ID, called a “seed”. Then, we use Kythe~\cite{web:blog:kythe} to find references to the seed in the entire Google codebase, called “direct references”. This process is repeated 3 times where references-to-references are also automatically found in a bread-first-search fashion, where each level is farther away from the initial seed.

The result of this Kythe search is a superset of files and lines that may potentially need to be modified. We filter this superset to identify the locations to be modified accurately, before the files are passed to our LLM for the actual modification. To improve accuracy, we use various pluggable and extensible strategies that help decide whether a specific location needs to be migrated. First, we pass the locations through regular expressions to identify if production code contains any casting in different languages, e.g. \texttt{(int) seed.getLong()}. Such locations are tagged as to-be-migrated. Then, for test code, we check whether values passed as IDs are larger than the 32-bit space, e.g. \texttt{seed.setLong(5L)}. When they fit the 32-bit space, they are marked as to-be-modified too, since we want to execute tests with values larger than the maximum 32-bit value.

The remaining locations are passed through more filters that eliminate false positives with additional techniques. One approach is to parse the source code to check whether a given location contains an actual call relevant to the seed or not. As an example, a method/function definition may have been found with the Kythe references search, but usually function definitions are not relevant places to be changed for test files. Such locations are marked as irrelevant.

At the end of these filters, the remaining locations are kept to be passed to our LLM for migration. The human involvement in the process decreased during the development of the workflow but still remained high due to the prevalence of ambiguous locations to change.

\subsubsection{Code migration flow}\label{sec:toolkit}

To generate and validate the code changes we leverage a version of the Gemini model that we fine-tuned on internal Google code and data.

Each migration requires as input:
\begin{itemize}
    \item A set of files and the locations of expected changes: path + line number in the file.
    \item One or two prompts that describe the change.
    \item {[Optional]} Few-shot examples to decide if a file actually needs migration.
\end{itemize}

\begin{figure}[h]
\centering
\includegraphics[width=0.9\linewidth]{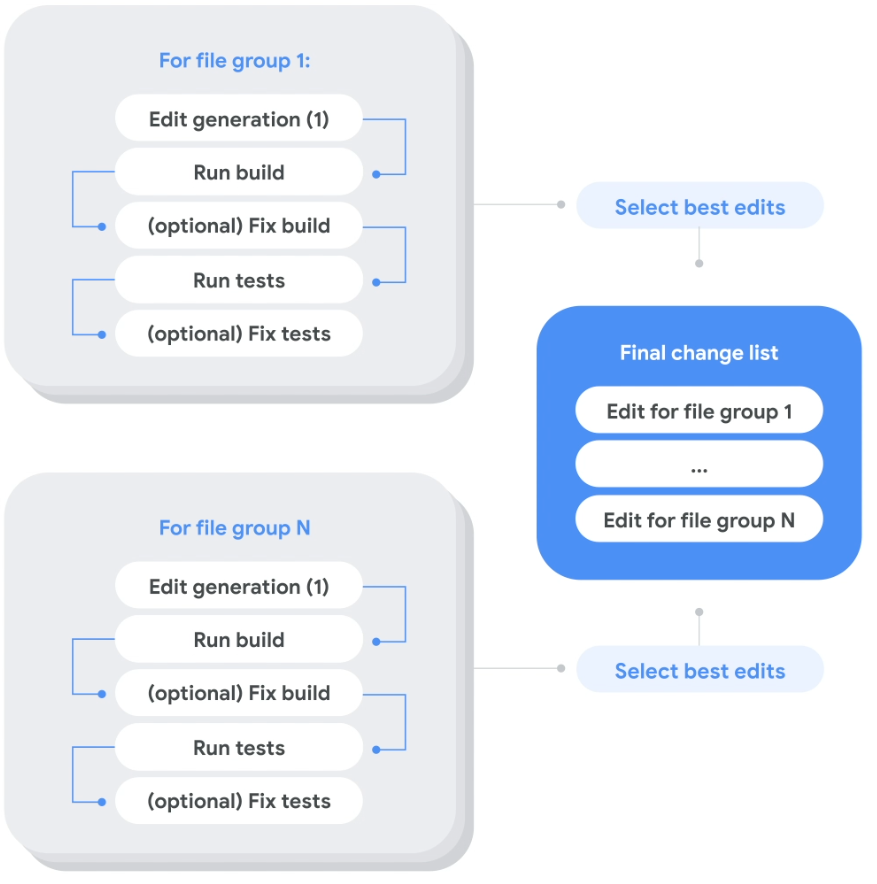}
\caption{Example execution of the multi-stage code migration process.}
\label{fig:mutistage}
\end{figure}

We have developed a code migration toolkit that is used in the solutions described in this work.  The toolkit is versatile and can be used for code migrations with varying requirements and outputs. All seed file change locations are provided by the user and collected through processes similar to the ones described above. The migration toolkit automatically expands this set with additional relevant files that can include: test files, interface files, and other dependencies. This step uses symbol cross-reference information.

In many cases, the set of files to migrate provided by the user is not perfect. It is not unusual for some files to have already been partially or completely migrated. Thus, to avoid redundant changes or confusing the model during edit generation we provide the model with few-shot examples and ask it to predict if a file needs to be migrated.

\begin{figure}[h]
\centering
\includegraphics[width=1\linewidth]{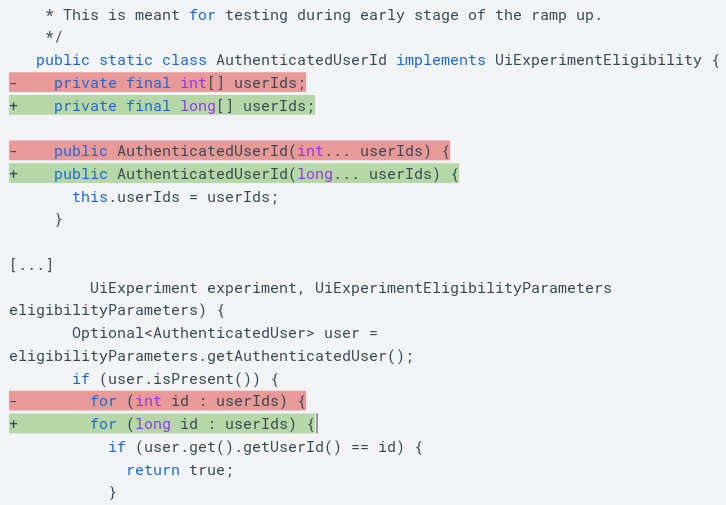}
\caption{In the example above we prompt the model to only update the constructor of the class where the type has to change. In the predicted unified diff, the model correctly also fixes the private field and usages within the class.}
\label{fig:code1}
\end{figure}

\begin{figure}[h]
\centering
\includegraphics[width=1\linewidth]{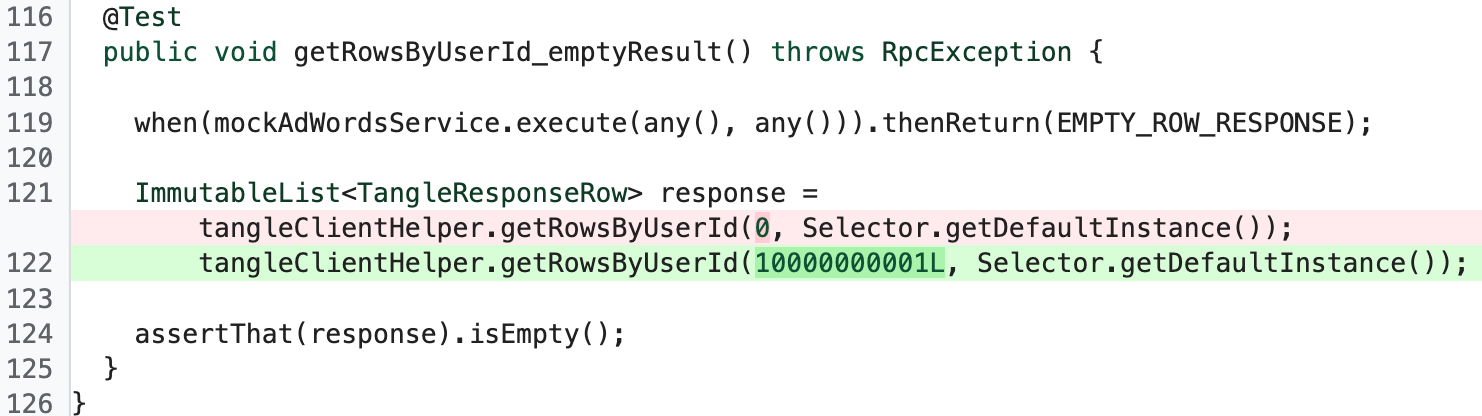}
\caption{The model updated also the test file with an integer that is larger than 32-bit.}
\label{fig:code_test}
\end{figure}

The edit generation and validation step is where we have found the most benefit from an ML-based approach. Our LLM was trained following the DIDACT \cite{web:blog:didact} methodology on data from Google’s monorepo and processes. At inference time, we annotate each line where we expect a change is needed with a natural language instruction as well as a general instruction for the model. In each model query, the input context contains one or multiple files related to each other, for example implementation files with headers, tests, interface declarations etc.. Fig~\ref{fig:mutistage} shows how related files are grouped together automatically and their combined changes are later added to the set of changes comprising a 'run' of the toolkit.  The model predicts differences (diffs) between the files where changes are needed and will also change related sections so that the final code is correct.
The instructions to the model are relatively simple (see Fig~\ref{fig:int32prompt}), but remind the model to also update the test files. Examples of consistent results can be seen in Fig~\ref{fig:code1} and Fig~\ref{fig:code_test}.

\begin{figure}

\begin{Verbatim}[samepage=false, commandchars=\\\{\}, fontsize=\scriptsize]
\{id\} should be of type int64_t.
Update the tests to reflect a large id.
Initialize the \{id\}s with values larger than 10000000000.
If necessary add new test parameters with large ids.
If previous id was negative, new value should be negative.
\end{Verbatim}

\caption{Prompt for int32 to int64 migration. The model makes the change across all the file(s) even without being asked.}
\label{fig:int32prompt}
\end{figure}

This last capability is critical to increase migration velocity, because the generated changes might not be aligned with the initial locations requested, but they will solve the intent. This reduces the need to manually find the full set of lines or files where changes are needed and is a big step forward compared to purely deterministic change generation based on AST modifications.

Different combinations of prompts yield different results depending on the input context. In some cases providing too many locations where one might expect a change results in worse performance than specifying a change in just one place in the file and prompting the model to apply the change to the file holistically. 

As we apply changes across dozens and potentially hundreds of files, we implement a mechanism that generates prompt combinations that are tried in parallel for each file group. This is similar to a pass@k \cite{chen2021evaluatinglargelanguagemodels} strategy where instead of just inference temperature we modify the prompting strategy.

We validate the resulting changes automatically. The validations are configurable and often depend on the migration. The two most common validations are building the changed files and running their unit tests. Each of the failed validation steps can optionally run an ML-powered “repair”. The model has also been trained on a large set of failed builds and tests paired with the diffs that then fixed them. For each of the build/test failures that we encounter, we prompt the model with the changed files, the build/test error and a prompt that requests a fix. With this approach, we observe that in a significant number of cases the model is able to fix the code.

As we generate multiple changes for each file group, we score them based on the validations and at the end decide which change to propagate back to the final change list (similar to a pull request in Git). There are different strategies to rank changes and select the best. Which to use depends on the use case, commonly a good approach is to provide examples of the expected diff and score the changes that are closer to the set of provided examples. Closer can be defined by standard code/text distance metrics or by querying Gemini asking which of the changes best matches the given examples.

The process and toolkit described in Sec~\ref{sec:toolkit} is mostly generic, and is under in the other migrations exercises that we present next.

\section{JUnit3 to JUnit4 migration}\label{sec:j3}
Large codebases tend to have some parts that begin to fall behind from the rest, they might use an older version of a library or API, or a framework that is being deprecated elsewhere. Google has had a one-version policy for years, which helps keep every dependency and library as fresh as possible. Nevertheless as standards change, some migrations that couldn't be fully automated tend to take quarters, even years.

A group of teams at Google had a substantial set of test files that used the now old JUnit3 library. Updating all of them manually is a huge investment and although the old tests are not on the critical path of the development, they negatively affect the codebase. They are technical debt and tend to replicate themselves, as developers might inadvertently copy old code to produce new one.

We needed to make a decisive push to migrate a critical mass of these tests to the new JUnit4 library. Although for a human such a migration is relatively simple, doing so with purely AST-based techniques was deemed infeasible as there are just too many edge cases.

\begin{figure}
\begin{Verbatim}[samepage=false, commandchars=\\\{\}, fontsize=\scriptsize]
You are a frontend software engineer that is an expert
at Java, JUnit3, and JUnit4.
Your work involves upgrading Java unit test files
from JUnit3 to JUnit4.
Convert all provided test files below
from JUnit3 to JUnit4.
Add imports for all the assert methods used
in this file. Here are some tips to keep in mind:

Steps:
1. Change the imports
 * Remove imports for anything under `junit.framework`
 or `junit.extensions`
 * Add `import static org.junit.Assert.*`
2. Remove the base class from the test.
JUnit4-style tests should not
extend `junit.framework.TestCase`
3. It is rarely necessary to have a base class
for JUnit4-style tests.
Often you can write custom rules to share code
4. Annotate the test class with `@RunWith(JUnit4.class)`
5. Annotate the test methods with `@Test`
...
\end{Verbatim}
\caption{Excerpt from the JUnit3 to JUnit4 prompt}
\label{fig:junitprompt}
\end{figure}

\begin{figure}[h]
\centering
\includegraphics[width=1\linewidth]{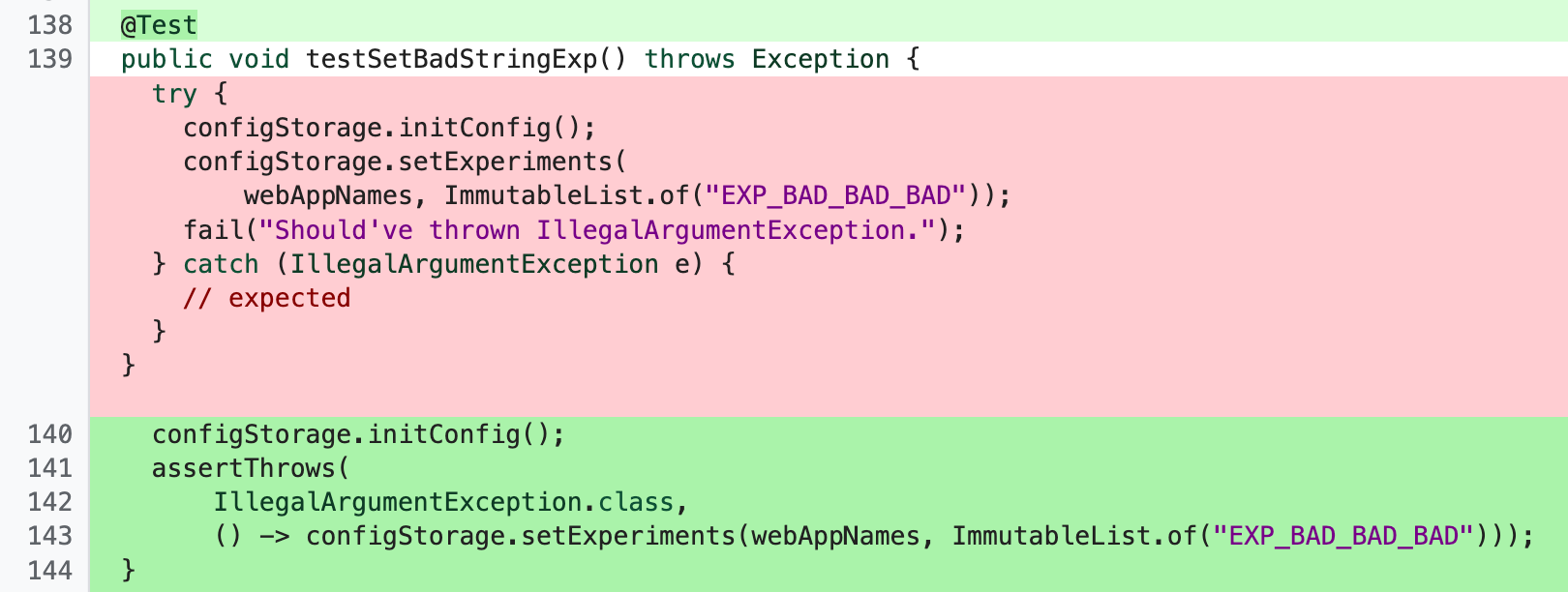}
\caption{Changes required for a case of JUnit3 to JUnit4 conversion.}
\label{fig:code2}
\end{figure}

We used the LLM migration stack (described above) to run an automatic migration on the old tests. We already had a list of all the JUnit3 tests in the repository, so running over all of them was simple. While we modified each test file, the internal system for Large Scale Changes (LSCs, \cite{winters2020software}) split the resulting changelists into smaller sets that were sent to the owners of the tests for review. An example change can be seen in Fig~\ref{fig:code2}.

The version of Gemini that we used was fine-tuned on the internal Google code base so it already had seen quite some JUnit4 tests. This allowed us to have a relatively simple set of prompts that essentially consisted of a list of rules (see Fig~\ref{fig:junitprompt}) that humans use to do the migration manually.

The updated test files were built and the updated tests ran again. Any failures were sent again to the model for fixing, as described in Fig ~\ref{fig:mutistage}.

With this technique we were able to migrate 5,359 files modifying more than 149,000 lines of code in 3 months. The bottleneck in the process was the speed at which engineers could review the changes. We purposefully limited the number of changes we generate every weak to avoid overwhelming reviewers.
At the end of the migration $\sim$87\% of the code generated by AI ended up committed without any change.

\section{Joda time to Java time migration}\label{sec:goodtime}

Some parts of the codebase still use the Joda time library for Java instead of the standard java.time package. Although the Java versions within the monorepo are regularly updated, this dependency remains and is still widespread with thousands of occurrences. We decided to tackle migrating from Joda time to standard java.time.

One major challenge in such a migration is that the changes are not scoped to singular methods but very often require changes in class public interfaces and fields. The situation becomes even more complex as we cannot just create a giant change and update all occurrences - there are thousands of them. Instead we need to split the work in chunks that can be reviewed and committed separately. The continuous availability of code reviewers is also not guaranteed all the time, so we need to only migrate parts of the codebase where we have bandwidth to land the changes. Often there are inter-dependencies between components we want to change and some other we do not want to touch now. This means we need to insert conversion functions in the interfaces where two such components interact and employ the types we want to migrate.

In many cases there is a 1:1 correspondence between the timekeeping functions and data structures, for example \texttt{\small joda.time.Duration} can be replaced with \texttt{\small java.time.Duration} and the constructing functions can be modified so that instead of \texttt{\small joda.time.Duration.millis()}, we call \texttt{\small java.time.Duration.ofMillis()}.

In other cases though there is no direct type translation possible. For example there is no counterpart to \texttt{\small joda.time.Interval} in the standard Java Time API. \texttt{\small Duration} has no concept of exact start time. Instead the guideline was to substitute a \texttt{\small joda.time.Interval} with a \texttt{\small common.collect.Range<java.time.Instant>}. This requires a more involved change in the logic of the functions and most importantly in the class interfaces that use this type.

All these challenges meant we need a new and more complex approach to gradually land the migration.
We split the process again in 3 stages: change targeting (also known as \emph{localization}), change execution, review and landing.

For the targeting we built a pipeline on top of Kythe \cite{web:blog:kythe} which provides cross-reference information. We start with the directories (which very roughly correspond to components and projects) where we have reviewers ready.
For each file in such a directory we build a cross-reference graph showing where the Joda time types are used and what the dependencies are.  We need answers to several questions:

\begin{enumerate}
    \item Which files should be migrated together?
    \item What is the fan-in/out of the call graph? A widely implemented interface might lead to a change in hundreds of files.
    \item Do the calls ‘escape’ the current component scope we have set for the set of changes? If yes, we might have to create transition wrappers to avoid ‘spreading’ the interface change outside.
\end{enumerate}

To answer these questions we built a clustering solution through Kythe where we categorize the potential changes. 
The cross-references form directed acyclic graphs (DAGs) connecting files. The simplest changes are in files where there are no modifications needed in an interface - this happens when the types to change are only used within the implementation of a class. The call-graphs tend to cluster and we split them into categories by number of files affected.

The clustering is important to also get the model to make consistent changes across files. When there is a dependency, ideally we migrate all files in one prompt. If we split them, Gemini effectively would not know, between inference invocations, if the referenced file is migrated or not - it needs to assume if a direct call is needed or a type conversion. The alternative is to show the already-migrated file to the model to avoid inconsistencies.
Fortunately, as Gemini offers a huge context window we can comfortably fit many of the clusters into the context window. This means we can prompt once for many files and get the whole cluster migrated. After the code change we run builds, tests and try to fix any eventual failures.

For the changes themselves, the instructions to the model are similar to the ones we provide human engineers. See Fig~\ref{fig:jodatime}.

\begin{figure}
\begin{Verbatim}[samepage=false, commandchars=\\\{\}, fontsize=\scriptsize]
Remove the usage of the joda.time classes
and instead use the standard java.time
module.
Update all usages of the Joda classes with
their standard counterparts. Import the
correct java.time module classes if needed.

Additional instructions:
* Instant is a drop-in replacement. So if
you see joda.time.Instant, you can easily
just replace it with a java.time.Instant.
* When you see something like new 
Instant(Long.MAX_VALUE), that is just 
java.time.Instant.MAX
* “new Instant(0)” on the other hand is
just java.time.Instant.EPOCH
* joda.time.Instant’s ctor’s replacement
is java.time.Instant.ofEpochMilli
* joda.time.Instant.getMillis() maps to
java.time.Instant.toEpochMilli()
* Don't use Instant.now() instead use
Timesource.system().now()
* joda.time.Duration can be replaced with
java.time.Duration.
* Instead of joda.time.Duration.millis(),
call java.time.Duration.ofMillis()
* Instead of joda.time.Duration.getMillis()
call java.time.Duration.toMillis()
* Instead of joda.time.Duration.standardSeconds()
call java.time.Duration.ofSeconds()
* Instead of joda.time.Duration.standardMinutes()
call java.time.Duration.ofMinutes()
* Prefer to use Instant over DateTime unless you
really need to print out or manipulate Dates vs
a specific point in time (eg Instant)
* If you really need joda.time.DateTime, use
java.time.ZonedDateTime as a replacement
* Don't use common.time.Clock. Use
common.time.TimeSource instead
* Interval is NOT a drop in replacement so
be very careful
* joda.time.Interval can be carefully
replaced by common.collect.Range<java.time.Instant>
* Caveat, joda.time.Interval is closed-open
so when creating the Range<Instant> and you
want 100% compatibility, you need a closed-open
Range.

Never rename functions or completely remove their
implementation.
\end{Verbatim}

\caption{Prompt for Joda Time conversion}
\label{fig:jodatime}
\end{figure}

\begin{figure}[h]
\centering
\includegraphics[width=1\linewidth]{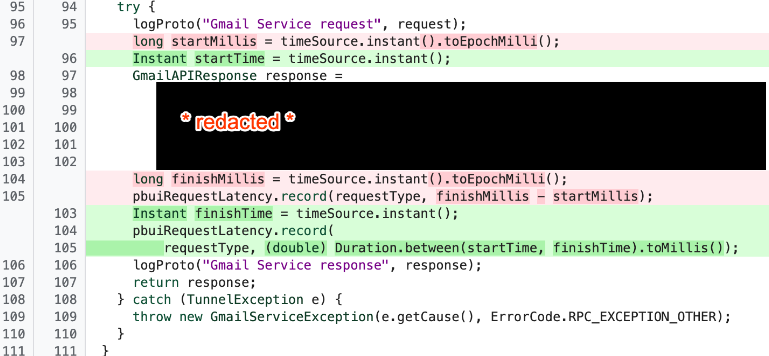}
\caption{Example of a Joda time to Java time migration}
\label{fig:code3}
\end{figure}

Gemini has seen enough Joda and java.time code during its training that we don’t actually have to explain what they are. It was able to correctly use the right APIs and didn’t hallucinate wrong parameters or data structures. Fig~\ref{fig:code3} shows an example of a correct diff generate for the migration.

The only additional context, apart from the files to migrate we offer, is an auxiliary class with conversion functions that sometimes need to be used. They are specific to the internal Google code and the model should consistently use them (instead of writing new ones).

This approach led us to successfully migrate many smaller and medium size file clusters. This migration is still ongoing, and we have challenges to solve. Sometimes, huge clusters arise and require us carefully ordering the changes and synchronizing with code reviewers. We need to improve some of the call-graph logic as connections are sometimes missed when components depend on each other through Guice\cite{web:guice}. Another challenge is the presence of switch statements that use dynamic typing. We are migrating our pipeline to support Guice dependencies though, which will cover the majority of the references.

Our current estimate shows that we are able to save $\sim$89\% of the time it would have taken humans to do the change in the small clusters. The number is calculated across multiple changes where human experts (team technical leads) have compared their experience with the AI-powered tooling to the previous purely-manual approach. An additional time-saver mentioned by engineers is that the current algorithm helps them quickly identify all places and dependencies to update. Even if some errors were made by the toolkit, they are relatively trivial to fix.

\section{Cleanup of Experimental Code}\label{sec:mendel}
Google uses thousands of runtime experiments running continuously to improve its products. At its most simplified, an experiment is a flag in code that receives some value from the outside (the experiment execution engine). The code logic is branched according to the value of the flag or is directly used somewhere as a parameter. After an experiment was either successful or not, the code related to it needs to be cleaned - either become the default code path or get removed entirely.

Sometimes, due to their large number, experiments become \textit{stale}: an experiment might become abandoned or permanently rolled out, or its value effectively becomes a constant, but the flag and the associated code remains even though no branching is needed anymore.
This constitutes dead code and technical debt. The experimentation engine tracks all experiments and flags across Google and can list all the stale flags - defined as flags whose value has not changed over a long period of time. Developers need to clean the code associated manually though.

Cleaning all these flags is time consuming and we decided to use AI to accomplish the job.

The task requires:
\begin{enumerate}
    \item Finding the code locations where the flag is referenced
    \item Deleting the code references to the experimental flag
    \item Simplifying any conditional expressions that depend on the experimental flag
    \item Cleaning any now dead code
    \item Updating the tests, often deleting now useless test
\end{enumerate}

Steps 2), 3), 4) could be implemented with an AST-based approach and even 5) could be approximated with a heuristic. In fact Google already had an AST-based tool for C++ but we discovered that it was missing some cases and the ambition was to build something that is language-agnostic.

\subsection{Flag discovery and targeting}

Step 1) is often simple - the flags are declared in configuration files and manifest themselves in code through getters whose symbol names are derived from the name of the flag in a well-known form. We use Code Search for finding where the flag is used in the code. One challenge is that in test files, the flag is not necessarily used with its exact getter name - in fact it is not extracted from the experiments system at all, but is a developer-written fake passed to the focal code. Software engineers use local variables with names that roughly align with the experiment flag name, but are not the same - in this case Code Search will not find the test flag.

\begin{figure}[h]
\centering
\includegraphics[width=1\linewidth]{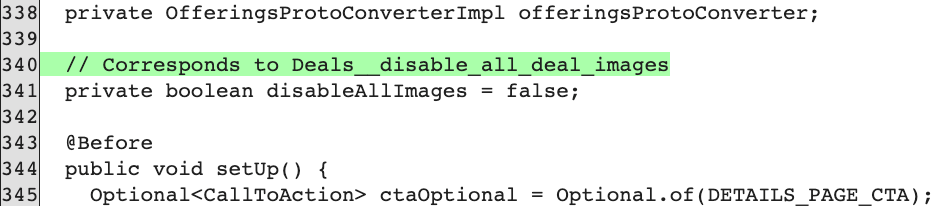}
\caption{The model discovers and 'tags' a test flag related to the implementation one we would want to delete}
\label{fig:code4}
\end{figure}

We could use cross-references information to try to map the uses of the experiment flag between the focal function and the test variable (often passed to a constructor). Human intuition however is enough to easily identify the test flag name when looking at the test and implementation files, so we decided to use the LLM to tell us which test flag we need to delete. Fig~\ref{fig:code4} illustrates how the LLM tagged it through adding a comment.  Through a Code Search query we know all the implementation files where the flag is used and due to the way tests are organized in Google, we also know all the corresponding test files. We pass all these files and the instructions below (see Fig~\ref{fig:mendel}) to the model to discover which test flags to delete.

\begin{figure}

\begin{scriptsize}
\begin{verbatim}
You are a software engineer trying to mark
an experiment parameter for colleagues.
Your main task is to add comments the Flag
in the test file that correspond to the
runtime usage of the flag in the
implementation file.
1. You are interested in the parameter
{PARAMETER_TO_DELETE} in the experiment
{EXPERIMENT_NAME}
2. In the test file add a comment with
the parameter {PARAMETER_TO_DELETE} name
to the test flags that corresponds to it.
3. The test flag that corresponds is one
that has a similar name AND is used in
methods where the original param is
expected.
4. For example a method that takes
{PARAMETER_TO_DELETE} in the implementation
file might take the flag to mark in the
test file.
5. Only add a comment in all lines where
the flag you identified to corresponds to
the parameter {PARAMETER_TO_DELETE} is
used in the Test file.
\end{verbatim}
\end{scriptsize}
\caption{Prompt for unused flag cleanup}
\label{fig:mendel}
\end{figure}

\begin{figure}[h]
\centering
\includegraphics[width=1\linewidth]{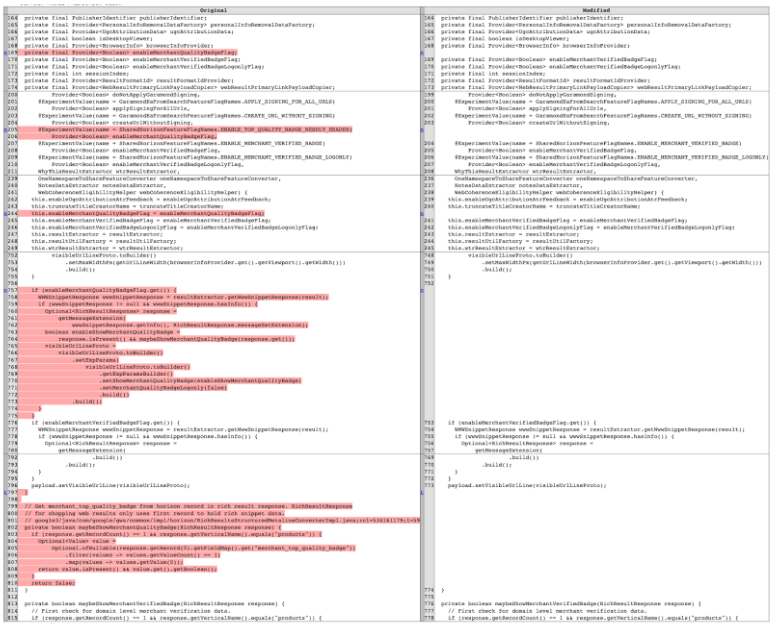}
\caption{The first large block with red background is a direct flag dependency, but the second large block with red background is dependency on removal of the first block.}
\label{fig:code5}
\end{figure}

Given the strong code change capabilities of Gemini, we decided to ask it to put a comment on the test flag. Then we use this data to guide the model in a second step that actually deletes the code across the implementation and test files (see Fig~\ref{fig:code5}).

\subsection{Code cleanup}

The code changes follow a similar pattern as the other migrations. The input to the model is:
\begin{itemize}
    \item Set of files and the symbol name of the flag to clean and the value of the flag
    \item Set of test files to clean the test flags to clean
    \item Instructions on how to execute the cleanup
\end{itemize}

The value of the flag is critical as we effectively need to substitute where it is used by that constant.

Given the large context size of Gemini we are able to pack all the usages of a flag in one query to the model, even though some files in the Google monorepo are very large. The implementation and test file are both visible together to the model so that it can make a consistent change across both of them.
After the code is cleaned we run additional validations to make sure all the instances were deleted, the code builds and tests pass.

The implementation files’ updates are of high quality and Gemini can successfully delete functions that are now dead code (because they were only used in a branch that is now deleted.)

The most challenging part is cleaning up the right tests. The model correctly deletes the tests where the flag is set to values that are now impossible. However in some cases the tests are not ‘pure’ and they might test multiple flags in the same unit test or some interaction between their values. Such cleanups are very difficult for humans as well. In these cases we rely on the test failing and the model attempting a fix or leave to the human reviewers to do a final fix. Addressing these unit tests cleanup is a future area of research for the team.

\section{Discussion and Takeaways}\label{learnings}

\subsection{LLMs for Code Modernization}
LLMs offer a significant opportunity for assisting modernizing and updating large codebases. They come with a lot of flexibility, and thus, a variety of code transformation tasks can be framed in a similar workflow and achieve success. This approach has the potential to radically change the way code is maintained in large enterprises. Not only can it accelerate the work of engineers, but make possible efforts that were previously infeasible due to the huge investment needed. Unfinished migrations tend to continuously slow down teams even if they are not working directly on them, they confuse new developers with obsolete patterns, and require additional cognitive load. Landing migrations faster has a wide reaching benefits beyond the actual technical code change. 

\subsection{LLM+AST, better together}
Many code migrations can be split into discrete steps and each step can either be LLM generation, LLM introspection, but also use traditional methods like AST-based techniques and even grep-like searches. We discovered that LLM planning capabilities are often \textit{not} needed and add a layer of complexity that should be avoided when possible. 

AST-based tooling has the advantage to be ‘always correct’ and not suffer from model version changes or prompt change fluctuations. For example simple mistakes of the model like adding unnecessary comments or changing methods it shouldn't, can easily be checked through an AST parser diffing the before/after of the file’s AST nodes. Working with smaller and better defined LLM-powered prompts increases the reliability of the results while reducing debugging efforts and the need to tune the prompts. An additional benefit is that the steps that don’t rely on LLMs tend to be much cheaper computation-wise. Although the cost per token for predictions has steadily decreased, migrations often require touching thousands of files and the costs might quickly add up.

\subsection{Divide and Conquer}
Simpler sub-tasks allow the developer building the migration workflow to iterate faster and ‘divide and conquer’ the problem. In the toolkit (see Sec~\ref{sec:toolkit}) described in the paper, different engineers are in charge of tuning and optimizing a specific step that is then used for multiple migrations. For example the build fixing step is ubiquitous and improving it leads to economies of scale.

When the model output is not perfect, leaning into validation and verification through multiple steps can ‘recover’ the change and lead to a successful workflow. In our observations, the LLMs are very good at fixing a well defined problem which complements their first attempt at a change.

\subsection{Landing the changes: a human process}
Although LLMs can lead to a significant time save for the change generation itself, additional tooling will be needed to further reduce or accelerate human involvement. Code reviews and change rollouts still require a human operator and can quickly become bottlenecks in the larger code migration process. We needed to slow down some of the migration work to avoid overwhelming the teams that had to do the reviews and to make sure that the rollouts happened in a gradual and responsible way.
Assisting the human developers in the tasks adjacent to the actual code changes in a migration is an area we are actively exploring.

\subsection{Metrics and evaluation}
A lot of work in the ML and applied ML communities focuses on evals, typically in which the model’s performance is measured on isolated, hermetic tasks such as “generate Python code to …”.  While these measure some aspects of model goodness, in the end, what matters is how model-level performance translates to business level outcomes. Bad model-level performance can compromise business-level outcomes, but good model-level performance does not guarantee good business-level outcomes, since the ML model is only one part of a complex transformation task.  We are constantly measuring ourselves at the level of business-level outcomes.

\subsection{Training and adoption}
The use of generative AI widely, with bespoke techniques, comes with a hidden cost: that of having to train a number of engineers in the use of these techniques.  Building elaborate tooling to completely hide the use of AI behind tools is expensive, and it creates a technical obligation to now maintain that tooling, which is used by a relatively small number of engineers.  (By comparison, generic technologies such as code auto-completion are easy to amortize over a much larger population.)

\subsection{Custom vs generic models}
Similarly to the above point, while fine-tuned models can be useful, they also come with a cost, and a company needs to constantly assess the “area under the curve” of investing in custom models for better outcomes, versus working with out of the box models.  We have currently used fine-tuned models for generating edits and fixing build failures.

\section{Related Work}

 \textbf{Repository-level changes with planning}: The need to be able to perform repository-level code changes is the key challenge of code migrations. For code migrations structural relationships across the repository are important LLMs to capture but the whole repository does not fit into the prompt. One approach discussed by Bairi et al. \cite{bairi2023codeplanrepositorylevelcodingusing} is framing the task as a multi-step planning problem, including a static dependency analysis, edit-relevance analysis (to determine the relevant context for the LLM), repairs on errors from validations in the previous step and adaptive plan generation. Jiang et al. presented in \cite{jiang2024selfplanningcodegenerationlarge} how LLMs can be used to generate the plan. Further relevant literature includes \cite{zhang2023planninglargelanguagemodels} employing look-ahead planning, \cite{ghallab2004automated} with a theoretical discussion. There are various demos of agent-based approaches \cite{web:Devin,yang2024sweagentagentcomputerinterfacesenable,zhang2024autocoderoverautonomousprogramimprovement}, e.g. giving the agent access to the editor, terminal and web. Although all these planning approaches showed promise to capture complex relationships and flexibly adapt during the execution, it is not clear how well the methods generalize and the complexity of the system is high.

\textbf{Repository-level changes without planning}: Xia et al. \cite{xia2024agentlessdemystifyingllmbasedsoftware} challenges above assumptions of planning being needed. In a simpler setup of a sequence of localizing the next relevant edit locations and repairing based on validation, better quality results are achieved on SWEBench \cite{web:swebench} compared to more involved approaches with planning. Also, in typical setup such an approach is more cost-effective. Similarly, approaches in this paper rely on agent-less techniques.

\textbf{Code migrations}: There is prior art in using LLMs for language translation \cite{eniser2024translatingrealworldcodellms} and code refactoring \cite{Shirafuji_2023}. However, those approaches do not reach high enough quality to become useful yet.  Amazon recently shared a product for code migrations using agents \cite{web:amazonQ}. In \cite{amazonQ} the team explores the human-AI partnership in the product: the  shortcomings of LLMs are attempted to be compensated by offering the human the ability to easily correct intermediate steps of code edits by the LLM.

\section{Conclusion and Future work}
We intend to build upon this success and expand the portfolio of programs leveraging LLM capabilities from a handful to several across multiple teams in Ads and other product areas across Google. Scaling AI-assisted migrations through easy-to-use workflows and high confidence AI recommendations will be critical to expanding adoption. 

We also plan to expand the portfolio of use cases from code migration to creating agents that can automate triaging, mitigating and resolving complex system escalations that will help the business run more smoothly with minimal interruptions.

\section{Acknowledgements}
We would like to thank our colleagues for work towards the above features and useful discussions: Max Kim, Kaitlin Barnes, Khushbu Patil, Gaoxiang Chen, Yurun Shen, Ayoub Kachkach.
 
%\bibliographystyle{IEEEtran}
%\bibliographystyle{IEEEtranN}
%\bibliography{references}

\printbibliography

\end{document}